\documentclass[12pt]{article}

\usepackage{graphics,epsfig}
\newcommand{\ci}[1]{\cite{#1}}
\newcommand{\bi}[1]{\bibitem{#1}}
\newcommand{\lab}[1]{\label{#1}}

\newcommand{\ba}{\begin{eqnarray}}
\newcommand{\ea}{\end{eqnarray}}
\newcommand{\beqs}{\begin{eqnarray}}
\newcommand{\eeqs}{\end{eqnarray}}

\begin{document}
\title{Gravitation  form-factors and spin asymmetries in hadron elastic scattering
}
\author{O. V. Selyugin\fnmsep\thanks{\email{selugin@theor.jinr.ru}  } \\
 BLTP, Joint Institute for Nuclear Research,
141980 Dubna, Moscow region, Russia 
}

\begin{center}
\textbf{Gravitation  form-factors and spin asymmetries in hadron elastic scattering}

\vspace{5mm}

 \underline{O.V. Selyugin}$^{\,1}$

\vspace{5mm}

\begin{small}
  (1) \emph{BLTPh, JINR, Dubna, Russia} \\
  $\dag$ \emph{E-mail: selugin@theor.jinr.ru}
\end{small}
\end{center}

\vspace{0.0mm} 

\abstract{
 In the framework of the model, where the scattering amplitude is determined
 by the first and second moments of the GPDs, the qualitative description of all
  existing experimental data at $\sqrt{s} \geq 52.8 $ GeV, including
  the Coulomb range and large momentum transfers, is obtained with only
  4 free parameters.  The spin-flip amplitude of the nucleon-nucleon elastic scattering
 is determined  taking into account the  spin-dependence part of the second moment of the generalized
 parton distributions (GPDs) with a new set of $t $-dependence.
   The corresponding value of the 
    $A_N$ for the $pp$ at
  high energy (RHIC) elastic scattering is obtained.
  }
%
\section{Introduction}

	The dynamics of strong interactions  finds its most
  complete representation in elastic scattering at small angles.
  Only in this region of interactions can we measure the basic properties that
  define the hadron structure: the total cross section,
  the slope of the diffraction peak and the parameter $\rho(s,t) $.
  Their values
  are connected on the one hand with the large-scale structure of hadrons and
  on the other hand with the first principles which lead to the
  theorems on the behavior of the scattering amplitudes at asymptotic
  energies \ci{mart,roy}.


       There are many different models for the description of hadron elastic
 scattering at small angles \cite{Rev-LHC}
They lead to  different
 predictions for the structure of the scattering amplitude at asymptotic
 energies, where the diffraction  processes can display complicated
 features \cite{dif04}.  This concerns especially the asymptotic unitarity
 bound connected with the Black Disk Limit (BDL) \cite{CPS-EPJ08}.
  saturation regime, which will be reached at the LHC
\cite{SelyuginBDLCJ04,CS7}. 
The effect of  saturation will be a change in the $t $-dependence of $B $ and $\rho $,
which will begin for $\sqrt{s} = 2 $ to  6  TeV, and which may drastically
change $B(t) $ and $\rho(t) $ at $\sqrt{s} = 14 $~TeV \cite{CSPL08}.
As we are about to see, such a feature can be obtained in very different models.

  Spin effect play very often the crushable role for the many different theoretical approaches.
     Now there are many different models for the description of the elastic
  hadrons scattering amplitude at small angles \cite{mog1}.
  They lead to  different predictions of the structure of the scattering
  amplitude at super-high energies.

  The total helicity amplitudes can be written as $\Phi_{i}(s,t) =
  \Phi^{h}_{i}(s,t)+\Phi^{\rm em}_{i}(s,t) e^{\varphi(s,t)} $\,, where
 $\Phi^{h}_{i}(s,t) $ comes from the strong interactions,
 $\Phi^{\rm em}_{i}(s,t) $ from the electromagnetic interactions and
 $\varphi(s,t) $
 is the interference phase factor between the electromagnetic and strong
 interactions \cite{selmp1,selmp2,PRD-Sum}. 

 In the  impact-parameter representation
  the  Born term of the scattering amplitude will be
 \begin{eqnarray}
 \chi(s,b) \  \sim 
   \ \int \ d^2 q \ e^{i \vec{b} \cdot \vec{q} } \  F_{\rm Born}\left(s,q^2\right)\,,
 \label{tot02}
 \end{eqnarray}
 where $t= -q^2 $ and dropping the kinematical factor $1/\sqrt{s(s-2m_p^2)} $
 and a factor $s $ in front of the scattering amplitude.
  After  unitarisation, the scattering amplitude becomes
  \begin{eqnarray}
 F(s,t)  \sim
    \ \int \ e^{i \vec{b} \vec{q} }  \ \Gamma(s,b)   \ d^2 b\,.
 \label{overlap}
 \end{eqnarray}
In this work the standard eikonal unitarisation scheme is used which leads to the
standard regime of saturation, {\it i.e.} the  BDL \cite{Bog100}:
     \begin{eqnarray}
  \Gamma(s,b)  = 1- \exp[-i\chi(s,b)] .
 \label{overlap}
\end{eqnarray}
The overlap function $\Gamma(s,b) $ can be a matrix,
 corresponding to the scattering of different spin states.
 Unitarity of the $S $-matrix, $SS^{+} \leq 1 $, requires that $\Gamma(s,b) \leq 1 $.
 There can be different unitarization schemes which map $\chi(s,b) $ to different regions
 of the unitarity circle \cite{CSPPRD09}.

 In different models  one can obtain various pictures of
the profile function based on different representations of the hadron structure.
     In the model \cite{ModA} we suppose that the elastic hadron scattering amplitude can be divided in
   two pieces. One is proportional to the electromagnetic form factor. It plays the most important
    role at small momentum transfer. The other piece is proportional to
    the matter distribution in the hadron
    and is dominant at large momentum transfer.

As in the soft-hard pomeron  model (EPSH)~\cite{CSPL08}, we take into account the contributions of the soft and hard pomerons.
The nucleon-nucleon  elastic scattering amplitude is
proportional to the electromagnetic hadrons form-factors and can be approximated
at small $t $  by
\begin{eqnarray}
 T(s,t) \ =  [\ k_{1} \ (s/s_0)^{\epsilon_1}
           e^{\alpha^{\prime}_1 \  t \ ln (s/s_0)}
         \ + \    \ k_{2} \ (s/s_0)^{\epsilon_2}
             e^{\alpha^{\prime}_2 \  t \ ln (s/s_0)} ]
   \ G_{em}^2(t),
\end{eqnarray}
where $k_1=4.47 $  and $k_2 = 0.005 $ are the coupling of the ``soft''
 and ``hard'' pomerons, and $\epsilon_1 =0.00728 $, $\alpha^{\prime}_1=0.3 $,
 and $\epsilon_2=0.45 $, $\alpha^{\prime}_2=0.10 $
 are the intercepts and the slopes of the two pomeron trajectories.
The normalization $s_0 $ will be dropped below and
 $s $ contains implicitly the phase factor $\exp(-i \pi/2) $.
  I shall examine only high-energy nucleon-nucleon scattering with
 $\sqrt{s} \geq 52.8 $~GeV. So, the contributions
  of reggeons and odderon will be neglected. This model  only includes crossing-symmetric scattering
  amplitudes.
The assumption about the hadron form-factors leads to the amplitude
      \begin{eqnarray}
    T(s,t)_{Born.} = h_1 (F^{s}_{Born} +
    F^{h}_{Born})G_{em}^2+ h_2 (F^{s}_{Born} + F^{h}_{Born})G_{grav.}^2,
    \end{eqnarray}
  A non-linear trajectory for the pomeron is suppoused and, as a first approximation, it is assumed
  that the coupling is proportional to the gravitational form factor and that
  both soft and hard terms in the $F_{Born}(s,t) $ have $\alpha^{'}=0 $ at large $t $.

\section{Hadron form factors}
  As was mentioned above,  all the form factors are obtained from the GPDs of the nucleon.
The electromagnetic form factors can be represented as first  moments of GPDs (\cite{Ji97,R97})
\ba
 F_{1}^q (t) = \int^{1}_{0} \ dx  \ {\cal{ H}}^{q} (x, t); \ \ \
 F_{2}^q (t) = \int^{1}_{0} \ dx \  {\cal{E}}^{q} (x,  t).
\ea
 Recently, there were many different proposals for the $t $ dependence of GPDs.
 We introduced a simple form for this
 $t $-dependence~\cite{STGPD}, based on the original Gaussian form corresponding to that
 of the wave function of the hadron. It satisfies the conditions of non-factorization,
 introduced by Radyushkin, and the Burkhardt condition on the power of $(1-x)^n $
 in the exponential form of the $t $-dependence. With this simple form
  we obtained a good description of the proton electromagnetic Sachs form factors.
  Using the isotopic invariance we obtained good descriptions of the neutron
  Sachs form factors without changing any  parameters.

 The Dirac elastic form factor can be written
\begin{eqnarray}
 G^2(t)= h_{fa} e^{d_1 \ t} \ + \  h_{fb} e^{d_2 \ t} \
 + h_{fc} e^{d_3 \ t}.   \lab{eff}
 \end{eqnarray}
with $h_{fa}=0.55 $, $h_{fb}=0.25 $, $h_{fa}=0.20 $, and
 $d_1=5.5 $, $d_2=4.1 $, $d_3=1.2 $.
    The exponential form 
    lets us calculate the hadron scattering amplitude
    in the impact parameter representation~\cite{CSPL08}.
 The model used the GPDs of nucleon to obtain the gravitational form factor of the nucleon in the
impact-parameter representation.
This form factor can be obtained from the second momentum of the GPDs. Taking instead of
the electromagnetic current $J^{\mu} $ the energy-momentum tensor $T_{\mu \nu} $ together
with a model of quark GPDs, one can obtain the gravitational form factor of fermions
\ba
\int^{1}_{-1} \ dx \ x [H(x,\Delta^2,\xi) \pm E(x,\Delta^2,\xi)]  = A_q(\Delta^2) \pm B_{q}(\Delta^2) .
\ea
 For $\xi=0 $ one has
\ba
\int^{1}_{0} \ dx \ x [{\cal{H}}(x,t) \pm {\cal{E}}(x,t)] = A_{q}(t) \pm B_{q}(t).
\ea
Calculations in the momentum-transfer representation show
that the second moment of the GPDs, corresponding to the gravitional form-factor, can be
represented in the dipole form
\begin{eqnarray}
A(t)=L^2/(1-t/L^2)^2  .\label{overlap}
 \end{eqnarray}
with the parameter $L^2=1.8 $~GeV $^2 $.
For the scattering amplitude, this leads to
\begin{eqnarray}
A(s,b)  = \frac{L^5 b^3}{48} K_{3}(Lb), \label{K3}
 \end{eqnarray}
where $K_{3}(Lb) $ is the MacDonald function of the third order.
To match both parts of the scattering amplitude,  the
second part is multiplied by a smooth correction function which depends on the impact parameter
\begin{eqnarray}
\psi(b) = (1+\sqrt{r_{1}^2+b^2}/\sqrt{r_{2}^{2}+b^2}).\label{K3}
 \end{eqnarray}

The model has only four free parameters, which are obtained from a fit to
the experimental data:
 $$h_1=1.09 \pm 0.004; \ \ \ h_2=1.57 \pm 0.006; r_{1}^2=1.57 \pm 0.02 GeV^{-2} ; \ \ \ r_{2}^{2}= 5.56 \pm 0.06 (GeV^{-2}. $$
It was used all the existing experimental data at $\sqrt{s} \geq 52.8 $~GeV,
including the whole Coulomb region and up to the largest momentum transfers experimentally accessible.
In the fitting procedure,  only statistical errors were taken into account.
The systematic errors were used as an additional common normalization of the experimental data
from a given experiment. As a result, one obtains $\sum \chi^2_i /N \simeq 3 $ where $N=924 $ is
the number of experimental points  \cite{modA}. If one sums the systematic and
statistical errors, the $\chi^2/N $ decreases, to $2 $.
Note that the parameters are energy-independent.
The energy dependence of the scattering amplitude is determined
only by the intercepts of the soft and hard pomerons.
\label{sec:figures}
\begin{figure}
\includegraphics[width=0.5\textwidth] {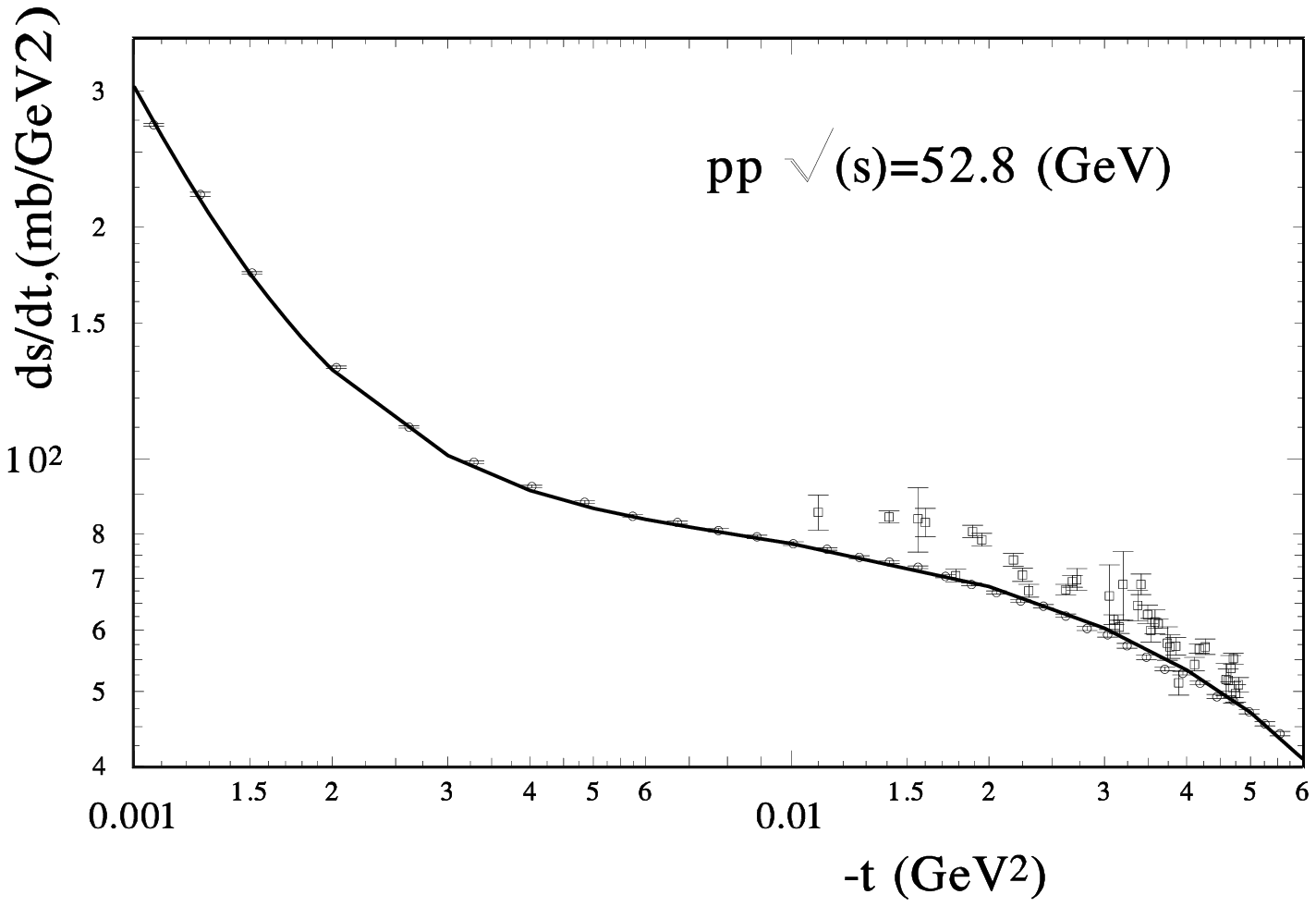}
\includegraphics[width=0.5\textwidth] {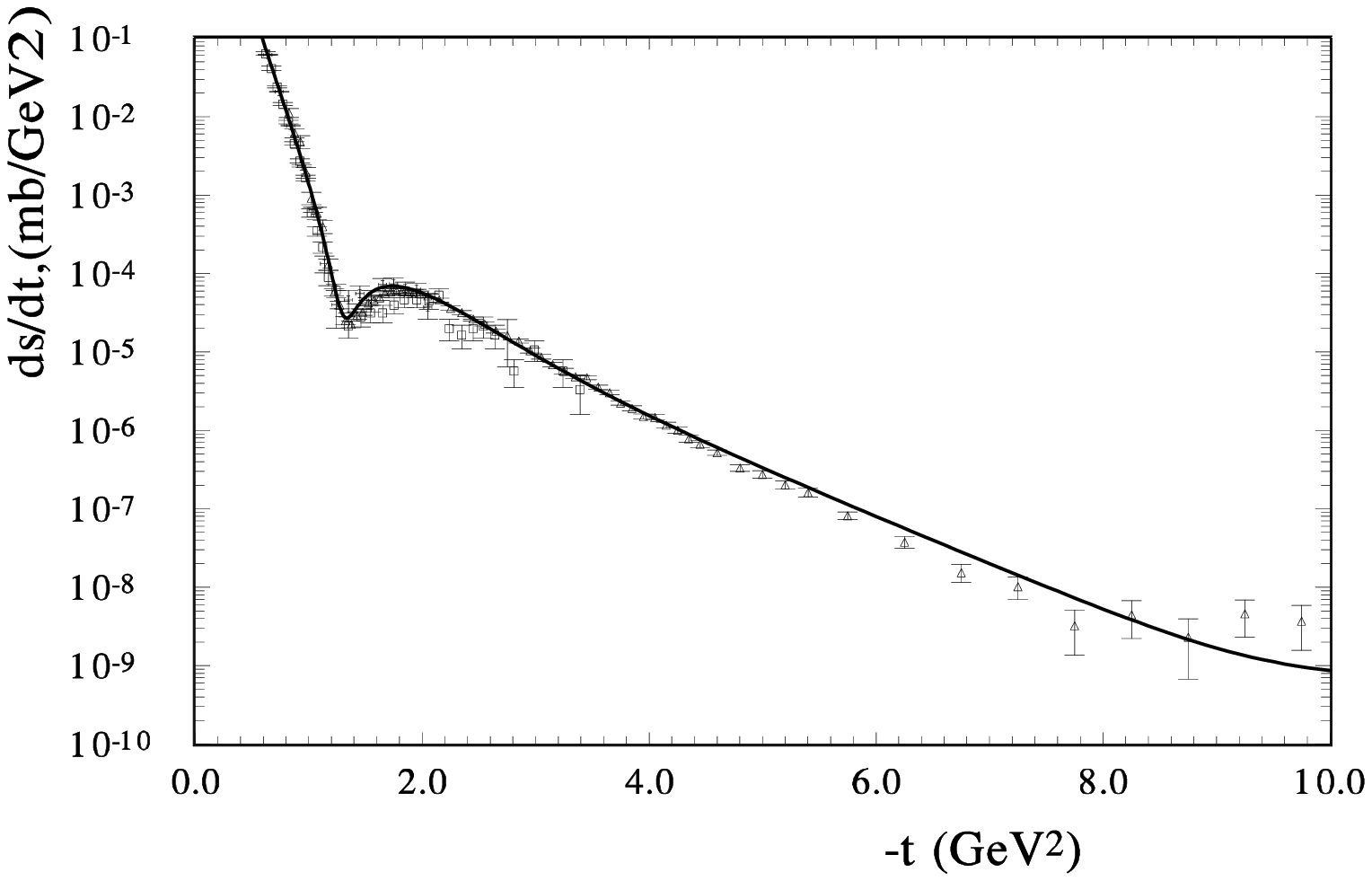}
\caption{ $d\sigma/dt \ {\rm at} \  \sqrt{s}=52.8 $~GeV for $pp $ elastic scattering,
at small $|t| $ (left) and at large $|t| $ (right). }\label{Fig:MV}
\end{figure}
In Fig.~1 the differential cross sections for proton-proton elastic scattering at
 $\sqrt{s} = 52.8 $~GeV
are presented. At this energy there are experimental data at small
(start at $-t=0.0004 $~GeV $^2 $) and large (up to $-t=10 $~GeV~$^2 $) momentum transfers.
The model reproduces both regions and provides a qualitative description of the dip region
at $-t \approx 1.4 $~GeV$^2 $, for $\sqrt{s}=53 $~GeV$^2 $ and for $\sqrt{s}=62.1 $~GeV$^2 $.

Let us examine the $p\bar{p}$ differential cross sections.
In this case at low momentum transfer
the Coulomb-hadron interference term plays an important role and has the opposite sign.
The model describes the experimental data well. In this case, the first part of the scattering
amplitude determines the differential cross sections, and is dominated by the exchange of the soft
pomeron. The high energy data at $\sqrt{s} = 630$ and $1800 \ $GeV also describe sufficiently well.
There was a significant difference between the experimental measurement of
 $\rho $, the ratio of the real part to the imaginary part of the scattering amplitude,
between the UA4 and UA4/2 collaborations at $\sqrt{s}=541 $~GeV.
A more careful analysis~\cite{SelyuginYF92,SelyuginPL94}  shows
that there is no contradiction between the measurements of UA4 and UA4/2.
Now the present model gives for this energy $\rho(\sqrt{s}=541 {\rm GeV}, t=0) = 0.163 $,
so, practically the same as in the previous phenomenological analysis.

Saturation of the profile function will surely control the behavior of $\sigma_{tot} $ at higher
energies and will result in a significant decrease of the LHC cross section.
For the last LHC energy $ \sqrt{s}=14 $~TeV the model predicts
 $\sigma_{tot}=146 $~mb and $\rho(0)=0.235 $. This result comes from the contribution of the hard pomeron
 and from the strong saturation from the black disk limit.

\section{Analysing power at high energies}

  There is a large spin program at  RHIC.
 This program includes measurements of the
 spin correlation parameters in the diffraction range of
 elastic proton-proton scattering.
The differential cross section and analyzing power $A_N$  are
defined as follows:
\begin{eqnarray}
\frac{d\sigma}{dt}&=& \frac{2
\pi}{s^2}(|\Phi_1|^2+|\Phi_2|^2+|\Phi_3|^2
  +|\Phi_4|^2+4|\Phi_5|^2), \label{dsth}\\
  A_N\frac{d\sigma}{dt}&=& -\frac{4\pi}{s^2}
                 Im[(\Phi_1+\Phi_2+\Phi_3-\Phi_4) \Phi_5^{*})],  \label{anth}
\end{eqnarray}
in terms of the usual helicity amplitudes $\Phi_i$.

In the general case, the  form of $A_N$ and the position of its
maximum depends on the parameters of the elastic scattering
amplitude: $\sigma_{\rm tot}$,  $\rho(s,t)$, the Coulomb-nuclear
interference phase  $\varphi_{\rm cn}(s,t)$ and the elastic slope
$B(s,t)$. For the definition of new effects at small angles, and
especially in the region of the diffraction minimum, one must know
the effects of the Coulomb-nuclear interference with sufficiently
high accuracy. The Coulomb-nuclear phase was calculated in the
entire diffraction domain taking into account  the form factors of
the  nucleons \cite{selmp1,selmp2,PRD-Sum}.

The total helicity amplitudes can be written as
\begin{eqnarray}
  \Phi_i(s,t) = \phi^h_{i}(s,t)
        + \phi_{i}^{\rm em}(t) \exp[i \alpha_{\rm em} \varphi_{\rm cn}(s,t)].
 \end{eqnarray}
In this paper, we define the hadronic and electromagnetic
spin-non-flip amplitudes as
\begin{eqnarray}
  F^{h}_{\rm nf}(s,t)
   &=&
    = \left[\phi^h_{1}(s,t) + \phi^h_{3}(s,t)\right]/2s; \ \ \ \
 F^{c}_{\rm nf}(s,t)
  = \left[\phi^{\rm em}_{1}(s,t) + \phi^{\rm em}_{3}(s,t)\right]/2s.
    \end{eqnarray}
 and spin-flip amplitudes as
\begin{eqnarray}
  F^{h}_{\rm sf}(s,t)
   &=&
    = \phi^h_{5}(s,t)/s ; \ \ \ \
 F^{c}_{\rm sf}(s,t)
  =  \phi^{\rm em}_{5}(s,t)/s.
    \end{eqnarray}
Equation (\ref{anth}) was applied at high energy and at small
momentum transfer, with the following usual assumptions
for hadron spin-flip amplitudes:
 $\phi_{1}=\phi_{3}$, $\phi_{2}=\phi_{4} = 0 \ $;
 the slopes of the hadronic spin-flip and spin-non-flip
amplitudes are equal.

  According to the standard opinion, the hadron spin-flip amplitude is
  connected with the quark exchange between the scattering hadrons,
  and at large energy and small angles it can be neglected.
  Some models, which take into account the non-perturbative effects,
  lead to the non-dying hadron spin-flip amplitude \cite{mog2}.
  Another complicated question is related with  the difference
  in phases of the spin-non-flip and spin-flip amplitude.

  Let us suppose that at high energies
   the spin-flip amplitude will be proportional the first and second momentum of the spin-depended part of the GPDs.
   \begin{eqnarray}
  F^{h}_{\rm sf}(s,t)
    = i \beta ( \int \ {\cal{E}}^{q} (x,  t) dx \  + \ \int \ x  \ {\cal{E}}^{q} (x,  t) dx).
    \end{eqnarray}
  Here we take only Born term of the spin flip amplitude, as it is small relative spin-non-flip
  amplitude and examine $A_N (s,t)$ 
  only at non-small $t$. 
  The coefficient $\beta$ is unknowing constant, which size $\beta=-0.01\sqrt{|t|}/m_p$ is determined by the size of the diffraction minimum   at the $\sqrt{s} = 52.8 \ $GeV.
    We do not know the relative sizes of the imaginary and real part of the spin-flip
    amplitude. For the simplicity in our model we take it as pure imaginary and without energy dependence.
    So, the ratio of the spin-flip to the spin-non-flip amplitude will be decreasing as $ln^{2}(s)$.
    Our calculations of the $A_N(s,t)$ was shown at Fig.2 and Fig.3. In Fig.2 the  calculation draw to
    low energy $\sqrt{s}=23.4  \ $GeV to compare with the existence the experimental data.
    Our model is essentially high energy approximation - it do not taking into account the contribution of the masses regions. However, the existence of the experimental data do not show the contradiction
    with our calculations. In fig.3 the predictions our model for the energies of RIHC are shown.
    The size of the $A_N$ for the maximal energy is not small and really can be measured.

\begin{figure}
\includegraphics[width=0.5\textwidth] {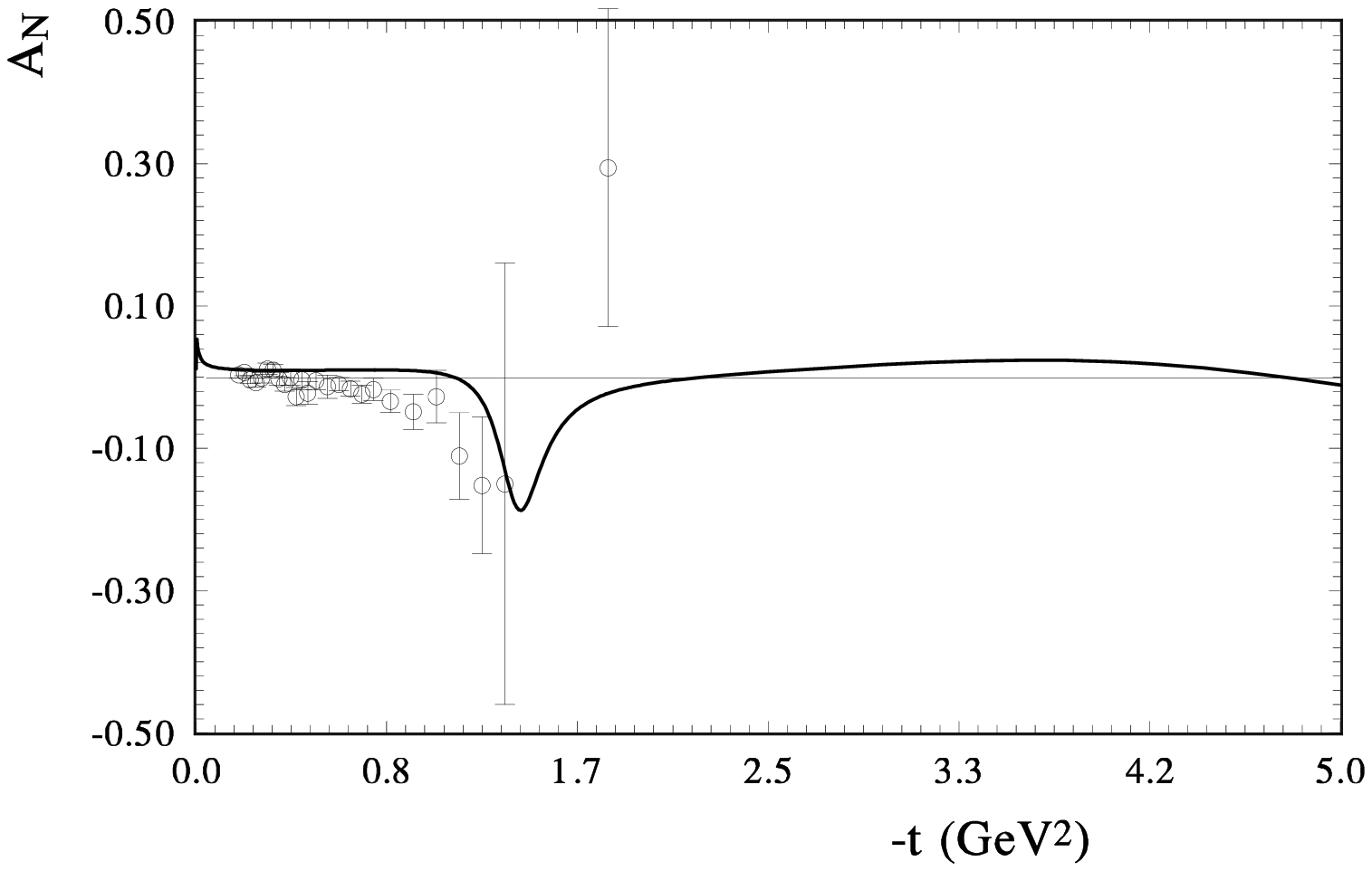}
\includegraphics[width=0.5\textwidth] {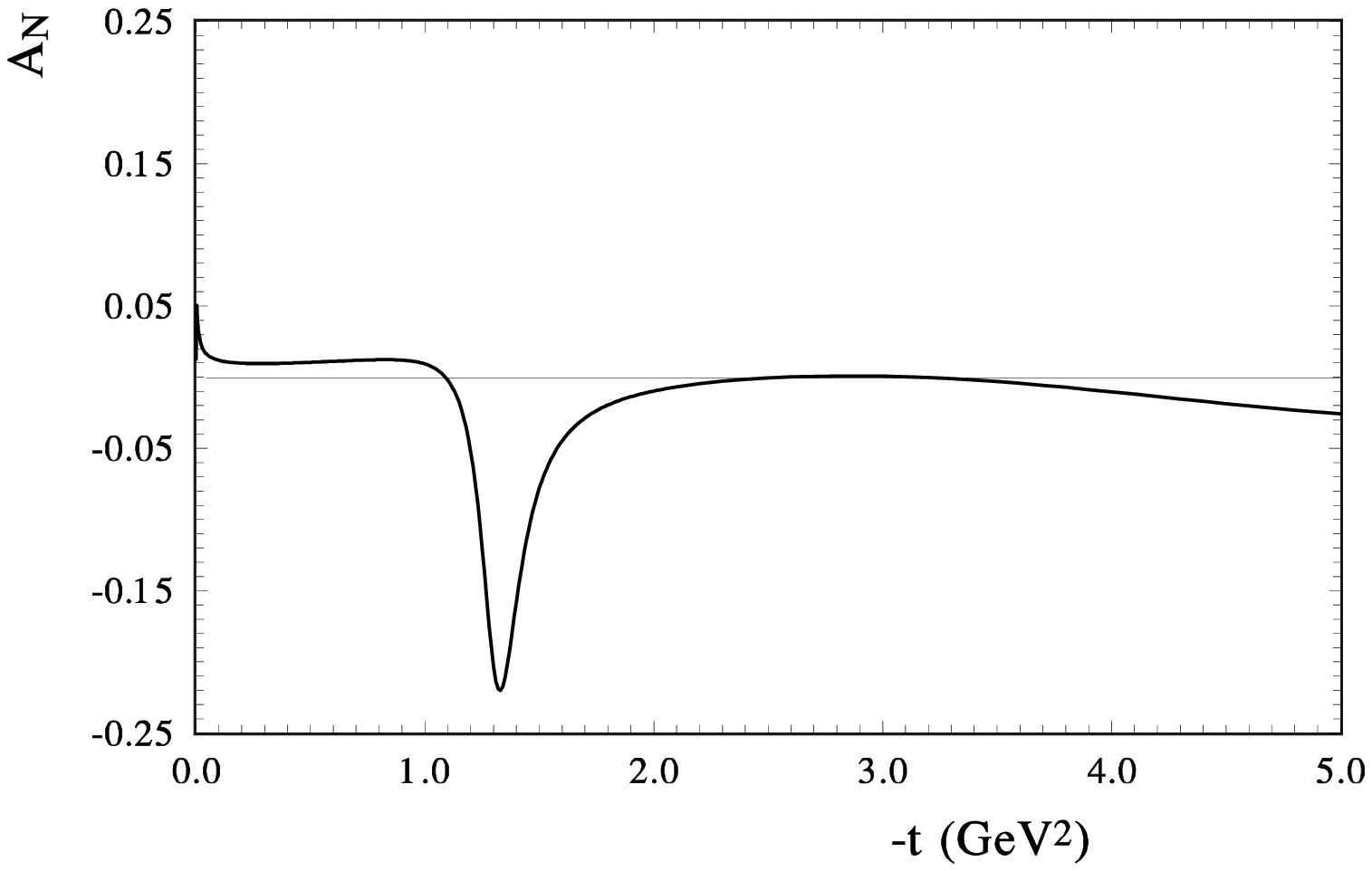}
 \caption{$A_N$ at $\sqrt{s} \ = \ 23.4 \ $\,GeV [left panel] and at $\sqrt{s} \ = \ 50 \ $\,GeV
 [right panel]; experimental points from 
  at  $\sqrt{s} \ = \ 23.4 \ $\,GeV.
 } \label{Fig_2}
\end{figure}

\begin{figure}
\vspace{-1.cm}
\includegraphics[width=0.5\textwidth] {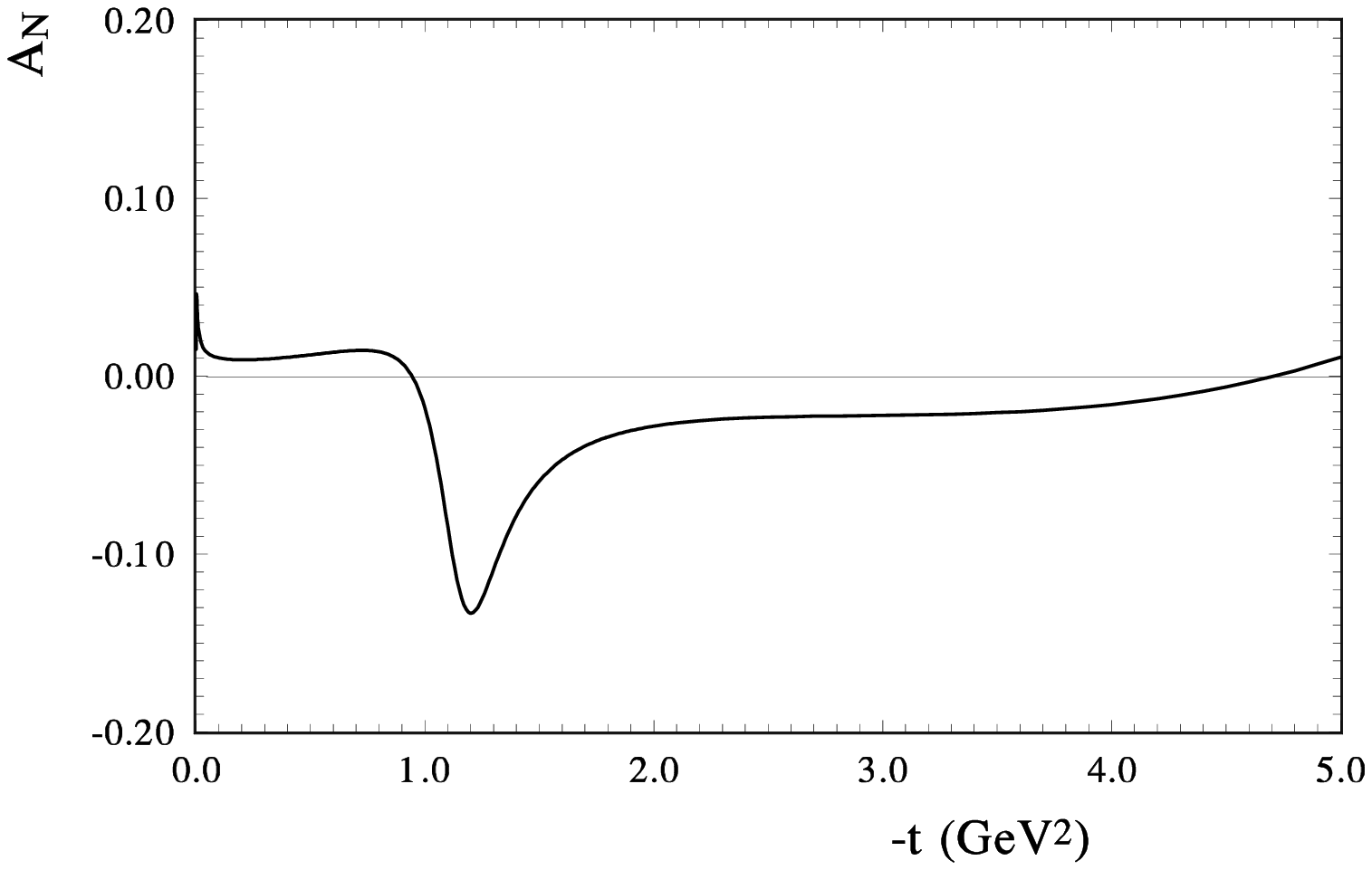}
\includegraphics[width=0.5\textwidth] {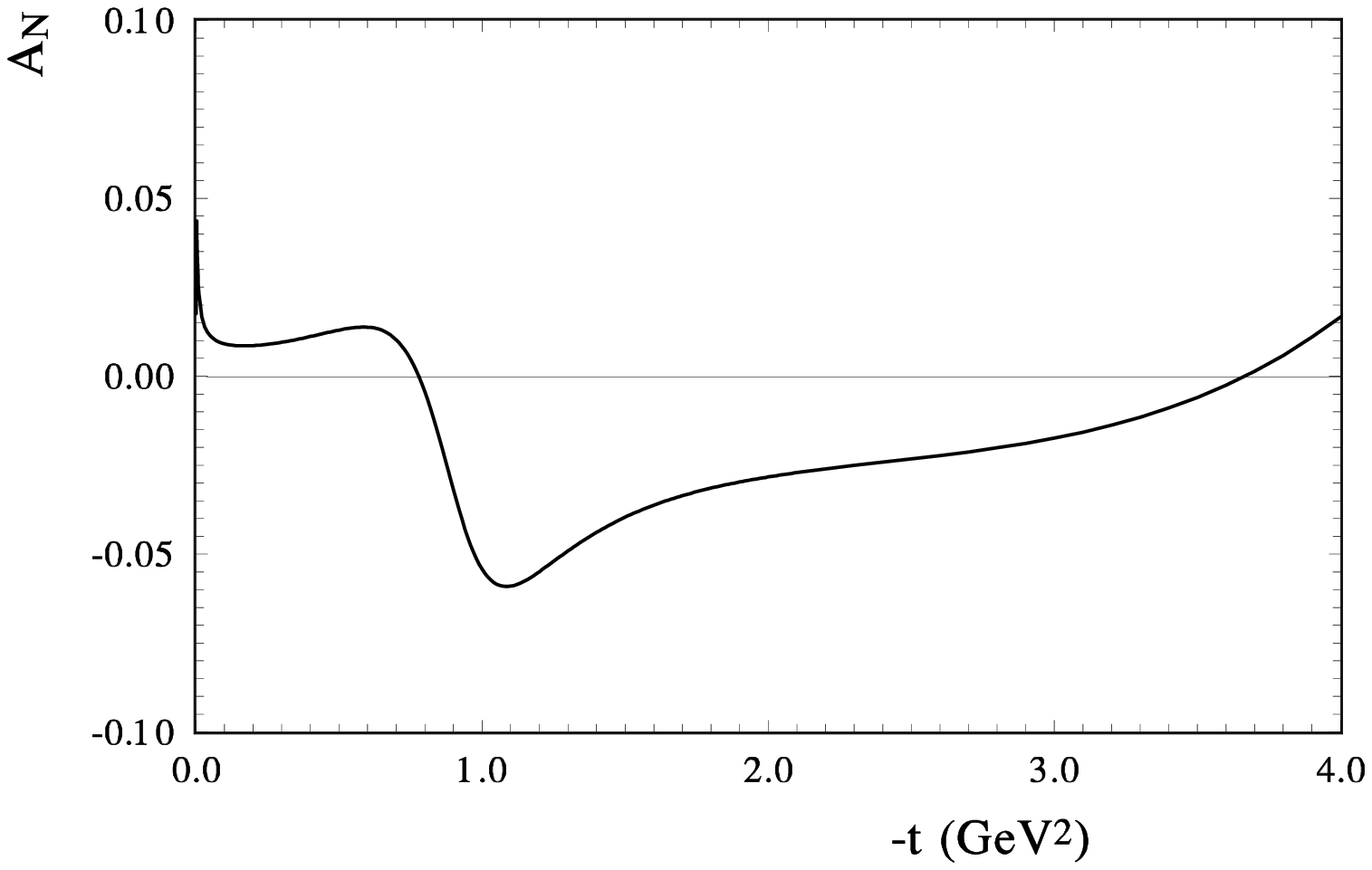}
 \caption{  $A_N$ at $\sqrt{s} \ = \ 200 \ $\,GeV  [left panel] and at $\sqrt{s} \ = \ 500 \ $\,GeV [right panel] calculated in the framework of the model.
 } \label{Fig_4}
\end{figure}

    \section{Conclusion}

A new model, taking into account the contributions of the
soft and hard pomerons and using form factors calculated from the GPDs, successfully describes all
the existing experimental data on elastic proton-proton and proton-antiproton
scattering at  $\sqrt{s} \geq 52.8 $~GeV, including the CNI region,
the dip region, and the large-momentum-transfer region. The behavior of the differential cross
section at small  $t $ is determined by the electromagnetic form factors, and at large  $t $ by the
matter distribution (calculated in the model from the second momentum of the GPDs).
 The spin flip amplitude which is determined by the spin-depended parts of the first and second momentum of
 the GPDs is calculated into the frame work of the model. It is need note that at large momentum transfer
 the spin-flip amplitude is determined in most part by the gravimagnetic second part of the GPDs $B(t)$.
 hence the measure of the analizing power at large momentum transfer and large energies can be help
 to determining its the parameters.
 The model is super simplified. Especially it is connect to the form of electromagnetic form factor,
     which is taken as the three exponents to calculate it's form by analitic in the impact
      parameter representation. The hard form factor also taking into account in the simple dipole form.
      The slope of the hard part of the Regge form of Pomeron taken as zero. It is means that
       really slope require the non-linear dependence of momentum transfer.
    It is need further develop of the model without changes the basis which gives the new view point
    on the hadron interactions at high energies.

\vspace{0.5cm}

{\small The authors would like to thank  for helpful discussions
 J.R. Cudell and
  gratefully acknowledges financial support
  from FRNS and  thank the  University of Li\`{e}ge
  where part of this work was done.
   }

\end{document}